\documentclass[a4paper, 12pt, fleqn]{article} 
\pdfoutput=1 

\usepackage[sc]{mathpazo}

\usepackage[T1]{fontenc}

\usepackage{iftex}
\ifpdf
    \usepackage[whole]{bxcjkjatype}
\fi

\newcommand{\tablesize}{\fontsize{10.5pt}{0.35cm}\selectfont}

\usepackage{titlesec}
\titleformat*{\section}{\fontsize{15pt}{\baselineskip}\bfseries}
\titleformat*{\subsection}{\fontsize{12pt}{\baselineskip}\bfseries\itshape}

\titlespacing*{\section}{0pt}{3.5ex plus 1ex minus .2ex}{2.3ex plus .2ex} 
\titlespacing*{\subsection}{0pt}{3.25ex plus 1ex minus .2ex}{1.5ex plus .2ex} 

\usepackage[top=20truemm,bottom=25truemm,left=20truemm,right=20truemm,footskip=13.5truemm]{geometry}

\usepackage[hyperfootnotes=false]{hyperref}

\hypersetup{hidelinks}

\usepackage{xcolor}

\usepackage{natbib}

\bibpunct{(}{)}{;}{a}{,}{;}
\usepackage{etoolbox}
\makeatletter
\pretocmd{\thebibliography}{\interlinepenalty=10000}{}{}
\makeatother

\usepackage{endnotes}
\let\footnote=\endnote
\makeatletter
\renewcommand\enoteformat{\rightskip\z@ \leftskip=1.8em \parindent\z@ \leavevmode\llap{\makeenmark\hspace{1.0em}}}

\makeatother

\usepackage{amsmath}

\usepackage{graphicx}

\usepackage{subcaption}

\usepackage{tikz}
\usetikzlibrary{shapes.geometric}
\usetikzlibrary{shapes.misc}

\usepackage{here}

\usepackage[labelfont=bf]{caption}

\usepackage{tabularx}

\newcolumntype{Y}{>{\centering\arraybackslash}X}
\newcolumntype{R}{>{\raggedleft\arraybackslash}X}
\newcolumntype{L}{>{\arraybackslash}X}

\newcommand{\startappendix}{
    \newpage
    \section*{Appendix}
    \setcounter{table}{0}
    \setcounter{figure}{0}
    \renewcommand{\thetable}{A\arabic{table}}
    \renewcommand{\thefigure}{A\arabic{figure}}
}

\title{\fontsize{17pt}{1cm}\selectfont{}Negative Impact of Online Political Incivility on Willingness to See Political Comments $^*$}
\author{Kohei Nishi $^\dagger$}
\date{}
\begin{document}
\maketitle
\vspace{-8mm}

\begin{center}\textbf{Abstract}\\\end{center}
\noindent
{\fontsize{10.5pt}{0.35cm}\selectfont
Recently, there has been significant attention on online political incivility. While previous research suggests that uncivil political comments lead people to be less willing to see more comments on the same issue, two critical questions have received limited exploration: (1) Are people exposed to uncivil political comments less willing to see other comments from the person who posted the uncivil comment?; (2) Are people exposed to uncivil political comments less willing to see comments from people who have different thoughts than them? To address these questions, the present study conducted a preregistered online survey experiment targeting Japanese citizens, focusing on the pro- vs anti-Kishida cabinet conflict in Japan. The results show that the participants were less willing to see other comments by the person who posted the comment when the comment was uncivil than when it was civil. In addition, the anti-Kishida participants were less willing to see political opinions posted online by people who have different thoughts than them when the comment was uncivil than when it was civil, while the participants in the other subgroups did not show a similar tendency. These findings suggest that uncivil expressions in online political communication might prompt people to avoid reading opinions from those who have different thoughts than them, which might promote political echo chambers.

\vspace{13mm}
\noindent\rule{2in}{0.5pt}

\noindent
$^*$ The first version of this working paper was uploaded on March 13, 2024, and this version on June 1, 2024. Note that it has not yet passed peer review.

\noindent
$^\dagger$ Research Fellow of Japan Society for the Promotion of Science and Ph.D. Student at the Division of Law and Political Science, the Graduate School of Law, Kobe University, Japan

}

\newpage
\section*{Introduction}
In recent years, researchers have shown an increased interest in the trend of political incivility \citep[e.g.,][]{Jamieson2015,Kenski2020,Muddiman2017,Papacharissi2004,Sobieraj2011,Theocharis2016}.\footnote{Political incivility is defined in the present study as ``a disrespectful or insulting expression that attacks an individual or group in political communication.''} Previous studies have revealed that online political incivility undermines political trust \citep{Borah2013}, fuels negative attitudes toward the discussion partners \citep{Hwang2018}, and strengthens perceived polarization \citep{Hwang2014}. These findings imply that online political incivility is one of the key concepts for understanding politics in the age of polarization. Amongst various studies on the effect of political incivility, one has shown that uncivil political comments lead people to be less willing to see more comments on the same issue \citep{Kim2019}. Such findings are important in that they suggest that uncivil political comments have impacts on people's attitudes about how they interact with political information, which might be important for understanding what drives people to biased information consumption (i.e., echo chamber phenomenon). To gain a more detailed understanding of the mechanism of the echo chamber phenomenon, the following questions should be considered: (1) Are people exposed to uncivil political comments less willing to see other comments from the person who posted the uncivil comment?; (2) Are people exposed to uncivil political comments less willing to see comments from people who have different thoughts than them?

To investigate these points, the present study conducted a preregistered online survey experiment with a randomized controlled trial (RCT) approach targeting Japanese citizens. In the experiment, civil or uncivil comments, posted online, regarding the evaluation of the Kishida cabinet in Japan were randomly presented to participants, and their subsequent responses were measured. The results of the statistical analysis show that the participants were less willing to see other comments by the person who posted the political comment when the comment was uncivil than when it was civil. In addition, participants in the anti-Kishida subgroup were less willing to see political opinions by people who had different thoughts than them when the comment was uncivil than when it was civil, while the participants in the other subgroups did not show a similar tendency. These findings suggest that uncivil expressions in online political communication might cause people to mute and unfollow those who post uncivil comments on social media platforms, and sometimes even avoid seeing opinions from those who have different thoughts than them. The two tendencies together might promote the formation of political echo chambers.

\section*{Hypotheses}
According to previous studies, online political incivility produces negative emotions in the people exposed to it \citep{Hwang2018,Kim2019}.
Considering the previous studies' finding that negative emotions such as disgust, anxiety, and anger play significant roles in shaping individuals' political attitudes or behaviors \citep[e.g.,][]{Clifford2019,Valentino2008,Wolak2022}, incivility, mediated by the negative emotions aroused by it, might affect their attitudes and behaviors in online communication platforms.

A study has shown that uncivil political comments lead people to be less willing to see more comments on the same issue \citep{Kim2019}. Another study (not focusing on political fields) suggests that those who post disrespectful content are more likely to be muted on social media \citep{Pena2014}. Similarly, other studies using panel surveys have found that those with higher perceived levels of exposure to uncivil political comments or hate speech posts tended to unfriend people with greater frequency on social media \citep{Goyanes2021,Kim2022,Lin2023}. Such evidence suggests that those who are exposed to uncivil comments tend to be less willing to see and read the related comments.

On the other hand, several studies suggest that exposure to online political incivility stimulates people to engage in online political discussions. Specifically, people exposed to uncivil political comments are more willing to take some actions online in response to the comments \citep{Roden2022}, post comments \citep{Ziegele2018}, and participate politically online \citep{Borah2014}. On the contrary, a comment thread regarding a news article without partisan cues is less likely to be engaged when the top comment on the thread is uncivil \citep{Lu2023}.

To sum up the above-mentioned studies, it is implied that uncivil political comments undermine people's willingness to see and read the related comments, while the effect of incivility on people's willingness to engage in political discussions (more actively than just reading comments) seems to be unclear due to mixed evidence. The present study primarily focuses on the former, the effect of uncivil political comments on willingness to see political comments, as investigating it could contribute to the understanding of the selective exposure and echo chamber phenomenon, which seems to be deeply related to political polarization. While there are several non-RCT studies regarding this issue, which are mentioned above \citep{Goyanes2021,Kim2022,Lin2023,Pena2014}, there is a limited number of RCT studies. One study that used RCT, Kim and Kim \citeyearpar{Kim2019}, found that uncivil political comments decrease individuals' willingness to read more comments on the same topic. While the contributions of the existing studies are noteworthy, the following two points remain unexplored: (1) Are people exposed to uncivil political comments less willing to see other comments from the person who posted the uncivil comment?; (2) Are people exposed to uncivil political comments less willing to see comments from people who have different thoughts than them?

Gervais \citeyearpar{Gervais2021} revealed that when people are exposed to uncivil political expression, they tend to feel the emotion of disgust. Similarly, Hwang et al. \citeyearpar{Hwang2018} found that exposure to political incivility induces moral indignation, which includes feelings of anger, disgust, and contempt. Based on the previous study's argument that interpersonal and moral disgust cause people to avoid the sources of the feelings \citep{Miceli2018}, those who are exposed to uncivil political comments may be driven to avoid similar content. In such avoidance behaviors, they may avoid not only content posted by the writer of the uncivil comment but also content by other people who hold opposing opinions by stereotyping them as uncivil. Therefore, the present study hypothesized as follows:

\newpage
\vskip\baselineskip
\noindent
\textbf{Hypothesis 1}: Individuals who are exposed to a political comment are less willing to see other comments from the person who posted the comment when the comment is uncivil than when it is civil.

\vskip\baselineskip
\noindent
\textbf{Hypothesis 2}: Individuals who are exposed to a political comment are less willing to see political comments from people who have different thoughts than them when the comment is uncivil than when it is civil.

\vskip\baselineskip
As the present study conducted an experiment using the context of the anti- vs pro-Kishida cabinet in Japan, these hypotheses were tested for each of the following four groups separately as preregistered: (a) anti-Kishida participants who are exposed to a pro-Kishida comment, (b) pro-Kishida participants who are exposed to an anti-Kishida comment, (c) participants with neutral attitudes toward the Kishida cabinet who are exposed to a pro-Kishida comment, and (d) participants with neutral attitudes toward the Kishida cabinet who are exposed to an anti-Kishida comment. Note that in case of pro- and anti-Kishida participants, the study showed them comments from the opposing camp and none from their own camp. This is because the fundamental intent of the present study was to examine how individuals respond to uncivil expression by those with opposing opinions.

\section*{Methods}
To test the hypotheses, the present study conducted an online survey experiment with an RCT approach, using the context of the anti- vs pro-Kishida cabinet conflict in Japan. The hypotheses and outline of the plans for the experiment were preregistered on the Open Science Framework repository (\url{https://osf.io/je73s?view_only=26016bbfb16248a1aba7f240011556a4}). After review and approval by the research ethics committee (IRB) of the author's institution,\footnote{Research Ethics Committee of the Graduate School of Law, Kobe University (ID: 050015)} the survey was conducted on February 10, 2024 (Japan Standard Time).

\subsection*{Sampling}
Japanese citizens aged 18 to 70 were recruited to participate via Lucid Marketplace. The target number of recruits was set to \textit{N} = 4,000.\footnote{The number of participants slightly exceeded the goal for technical reasons (\textit{N} = 12). The exceeded responses were excluded from the dataset in the statistical analysis to align the sample size with preregistered one.} A quota-sampling approach based on age and gender was employed. Recruited participants were directed to Qualtrics, a web-based survey platform, wherein the survey was implemented. The responses that were automatically judged as Speeders by Lucid Marketplace were not counted as completed responses in Lucid's system, and they were excluded from the dataset in the following statistical analysis.

\subsection*{Pre-treatment Measurement}
Firstly, participants were asked about their level of online media use and political interest. Subsequently, to measure their feelings toward the Kishida cabinet, they were asked, ``How much favorability or antipathy do you have toward the Kishida cabinet?''\footnote{The survey was conducted in Japanese. The survey questions presented in this paper in English are translations.} Participants answered using a seven-point scale: 1 = ``Strong antipathy,'' 2 = ``Antipathy,'' 3 = ``Rather antipathy,'' 4 = ``Neither favorability nor antipathy,'' 5 = ``Rather favorability,'' 6 = ``Favorability,'' 7 = ``Strong favorability,'' and ``Don't know/Don't answer.'' Those who answered 1 = ``Strong antipathy,'' 2 = ``Antipathy,'' or 3 = ``Rather antipathy'' on the question were anti-Kishida, while those who answered 5 = ``Rather favorability,'' 6 = ``Favorability,'' or 7 = ``Strong favorability'' were pro-Kishida. In addition, those who answered 4 = ``Neither favorability nor antipathy'' were considered to have a neutral attitude toward the Kishida cabinet.

\subsection*{Treatment}
Each participant was presented with one of four fictional comments regarding the Kishida cabinet in Japan: (1) an uncivil comment from an anti-Kishida individual, (2) a civil comment from an anti-Kishida individual, (3) an uncivil comment from a pro-Kishida individual, and (4) a civil comment from a pro-Kishida individual (see Table \ref{table_comments} for details).\footnote{Because of ethical considerations, immediately before the treatment, participants were informed that a comment criticizing those who have a certain opinion would be presented to them, and only those who agreed to this were allowed to proceed.} Anti-Kishida participants were presented with one randomly selected comment out of (3) and (4), while pro-Kishida participants were presented with one randomly selected comment out of (1) and (2). The other participants were presented with one randomly selected comment out of (1), (2), (3), and (4).

\begin{table}[ht]
    \caption{Comments Presented in the Experiment}
    \label{table_comments}
    \centering\tablesize
    \begin{tabularx}{170mm}{lLLL}
        \hline
         & Type & Comment (Original) & Comment (Translation)\\
        \hline
        (1) & an uncivil comment from an anti-Kishida individual & ``岸田内閣を称賛してるヤツらは、アホ丸出しなんだよ！'' & ``People who are praising the Kishida cabinet reveal themselves as idiots!''\\
        (2) & a civil comment from an anti-Kishida individual & ``岸田内閣を称賛している人たちは、よく考え直すべきだと思う。'' & ``People who are praising the Kishida cabinet should reconsider carefully.''\\
        (3) & an uncivil comment from a pro-Kishida individual & ``岸田内閣を批判してるヤツらは、アホ丸出しなんだよ！'' & ``People who are criticizing the Kishida cabinet reveal themselves as idiots!''\\
        (4) & a civil comment from a pro-Kishida individual & ``岸田内閣を批判している人たちは、よく考え直すべきだと思う。'' & ``People who are criticizing the Kishida cabinet should reconsider carefully.''\\
        \hline
    \end{tabularx}
    \centering\tablesize
    \begin{tabularx}{170mm}{L}
    Note: The following text was displayed just above the comment: ``On social media and comment sections on news sites, various political opinions are posted. For example, a comment as follows is posted.'' In addition, the following text was displayed just below the comment: ``Please answer the questions below considering such a situation.''
    \end{tabularx}
\end{table}

\subsection*{Post-treatment Measurement}
As a manipulation check, to measure the level of incivility of the presented comment as perceived by participants, they were asked, ``Is the above comment an uncivil expression against others?'' Participants answered using a seven-point scale: 1 = ``I do not think so at all,'' 2 = ``I do not think so,'' 3 = ``I somewhat do not think so,'' 4 = ``Neither,'' 5 = ``I somewhat think so,'' 6 = ``I think so,'' 7 = ``I strongly think so,'' and ``Don't know/Don't answer.''

To measure their willingness to see other comments from the person who posted the presented comment, participants were asked, ``Would you like to see other posts from the person who posted the above comment?'' They answered using a seven-point scale: 1 = ``I would not like to see them at all,'' 2 = ``I would not like to see them,'' 3 = ``I rather would not like to see them,'' 4 = ``Neither,'' 5 = ``I rather would like to see them,'' 6 = ``I would like to see them,'' 7 = ``I would like to see them very much,'' and ``Don't know/Don't answer.''

To measure their willingness to see political comments from people who have different thoughts than them, participants were asked, ``Would you like to see political comments from people who have different thoughts than you on the internet?'' They answered using a seven-point scale: 1 = ``I would not like to see them at all,'' 2 = ``I would not like to see them,'' 3 = ``I rather would not like to see them,'' 4 = ``Neither,'' 5 = ``I rather would like to see them,'' 6 = ``I would like to see them,'' 7 = ``I would like to see them very much,'' and ``Don't know/Don't answer.''

In addition, they were asked about their gender, age, education level, and household income.

After all the questions, a debriefing was conducted. More specifically, participants were informed that the presented comment was a fictional one that was created for the survey, and asked to consent about the usage of their response data. The responses of those who disagreed at this point were not used in the following statistical analysis (\textit{N} = 113 out of 4,000).\footnote{All participants whose responses were used in the statistical analysis agreed with the use of their data after they were debriefed. However, owing to an error by the researcher, some participants who did not complete the survey (a small portion of the participants) were not provided with debriefing information. That is, once the number of responses reached the required goal, further responses were terminated, and thus, the participants thereafter were not provided with the debriefing session. Only a small number of participants belonged to this condition, and their responses were not used in the statistical analysis. The author reported this unanticipated issue to the research ethics committee (IRB) of the author's institution.}

\subsection*{Statistical Analysis}
The means of the outcomes were computed for each of the two conditions (uncivil and civil comment conditions), and the differences between them were computed.\footnote{The statistical analysis was conducted with R version 4.4.0 \citep{R2024}. The tidyverse package \citep{Wickham2019} in R was used for data wrangling and visualization.} In addition, two-tailed Welch's \textit{t}-tests were conducted on the mean differences in the outcomes between the two conditions at the .05 level of significance. The hypotheses were tested for each of the following four groups separately as preregistered: (a) anti-Kishida participants exposed to a pro-Kishida comment, (b) pro-Kishida participants exposed to an anti-Kishida comment, (c) participants with a neutral attitude toward the Kishida cabinet exposed to a pro-Kishida comment, and (d) participants with a neutral attitude toward the Kishida cabinet exposed to an anti-Kishida comment.

In the statistical analysis, the responses of those with duplicated IP addresses were excluded from the dataset (\textit{N} = 2 out of 3,887).\footnote{These are very rare cases that happened for unknown reasons, given that Lucid rejects participation with non-unique IP addresses by default.} In addition, the responses of those who failed an attention check question were excluded from the dataset (\textit{N} = 189 out of 3,885).\footnote{As an attention check, each participant was randomly presented with one of the following five sentences: ``Please select `Strongly disagree' for this question,'' ``Please select `Disagree' for this question,'' ``Please select `Neither agree nor disagree' for this question,'' ``Please select `Agree' for this question,'' and ``Please select `Strongly agree' for this question.'' The participants who did not select the proper item for this question were considered satisficers because they were considered not to have carefully read the question text.} Those who answered inappropriate ages (\textit{N} = 471 out of 3,696) were also excluded.\footnote{More specifically, if a respondent's answer to the age question on Lucid Marketplace and that on Qualtrics differed by more than one, the responses were judged as inappropriate. A difference of $\pm$ 1 year was accepted. Lucid Marketplace stores and reuses their age responses for 30 days (see \url{https://support.cint.com/s/article/Collecting-Data-From-Redirects}). Some of the participants might have had birthdays in the past 30 days and their ages have changed. Also, some may not remember their age exactly, and thus, the present study considered that an error of $\pm$ 1 year is natural and acceptable.} \footnote{The setting in which participants from the pilot studies were not allowed to participate in the main survey was preferable. Therefore, it was set up so that those who completed the pilot studies could not participate in the main survey. However, owing to an error by the researcher, those who participated in the pilot studies but did not complete them were able to participate in the main survey. Therefore, based on the panelist ID (PID) provided by Lucid, the response of a participant in the main survey who had also participated in the pilot study and seen the treatment comment was excluded from the dataset (\textit{N} = 1).}

The missing values in the outcome variables (``Don't know/Don't answer'') were processed by multiple imputation method with the Expectation-Maximization with Bootstrapping (EMB) algorithm \citep{Honaker2011}.\footnote{The present study generated \textit{M} = 1,000 imputed datasets using the Amelia II package \citep{Honaker2011} in R, analyzed the imputed datasets separately, and then combined the results using mi.t.test() function in the MKmisc package \citep{Kohl2022} in R. A model for predicting missing values included several variables, such as the experimental condition (as represented by a dummy indicator for the uncivil comment condition), age, gender identity, education level, and income level.}

\section*{Results}
\vspace{-7mm}
\subsection*{Results for Hypothesis 1}
\begin{figure}[ht]
    \centering
    \includegraphics[scale=0.4]{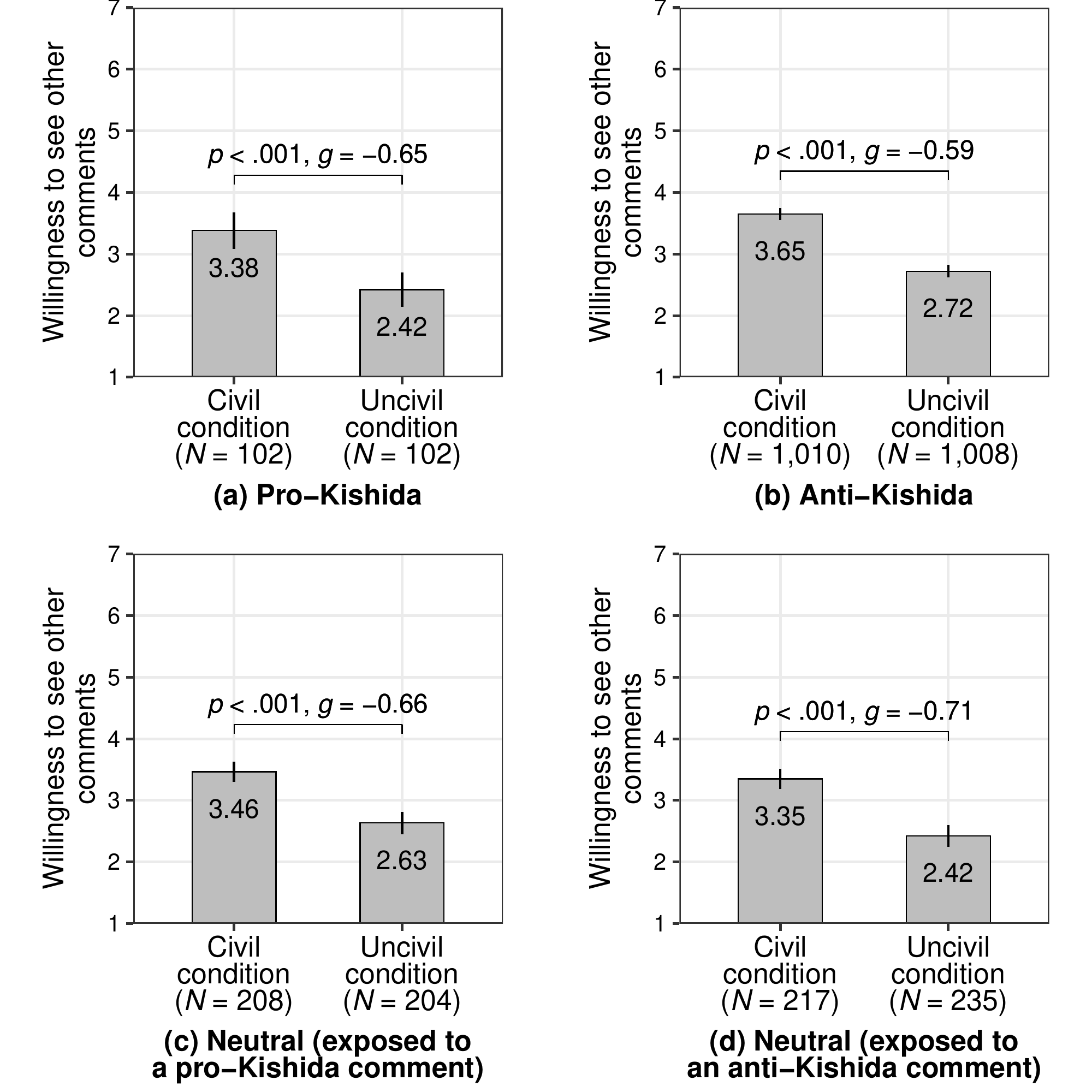}
    \caption{Results for Hypothesis 1}
    \label{figure_h1}
\end{figure}

Figure \ref{figure_h1} displays the results for Hypothesis 1. The sizes of the bars represent the mean values of the outcome variable, and the lines on the bars are their 95\% confidence intervals.

Consistent with Hypothesis 1, the participants were less willing to see other comments by the person who posted the comment when the comment was uncivil than when it was civil.
More specifically, for pro-Kishida participants (panel (a); upper left), the mean willingness of the participants exposed to a civil comment was 3.38 with 95\% CI [3.08, 3.68] on a 7-point scale, while the mean of those exposed to an uncivil comment was 2.42 with 95\% CI [2.14, 2.70]. The mean difference between the conditions was -0.96 (Hedges' \textit{g} = -0.65), which was statistically significant (\textit{t}(199.12) = -4.68, \textit{p} < .001). 
For anti-Kishida participants (panel (b); upper right), the mean willingness of the participants exposed to a civil comment was 3.65 with 95\% CI [3.55, 3.75], while the mean of those exposed to an uncivil comment was 2.72 with 95\% CI [2.62, 2.82]. The mean difference was -0.93 (Hedges' \textit{g} = -0.59), which was statistically significant (\textit{t}(1,988.09) = -13.09, \textit{p} < .001). 
For neutral participants exposed to a pro-Kishida comment (panel (c); lower left), the mean willingness of the participants exposed to a civil comment was 3.46 with 95\% CI [3.30, 3.63], while the mean of those exposed to an uncivil comment was 2.63 with 95\% CI [2.45, 2.82]. The mean difference was -0.83 (Hedges' \textit{g} = -0.66), which was statistically significant (\textit{t}(399.59) = -6.65, \textit{p} < .001). 
For neutral participants exposed to an anti-Kishida comment (panel (d); lower right), the mean willingness of the participants exposed to a civil comment was 3.35 with 95\% CI [3.18, 3.52], while the mean of those exposed to an uncivil comment was 2.42 with 95\% CI [2.25, 2.60]. The mean difference was -0.93 (Hedges' \textit{g} = -0.71), which was statistically significant (\textit{t}(434.54) = -7.51, \textit{p} < .001).

The differences are statistically significant in each of the four subgroups even with multiple testing correction with the Benjamini-Hochberg method \citep{Benjamini1995}, and thus, the findings support Hypothesis 1.

\subsection*{Results for Hypothesis 2}
\begin{figure}[ht]
    \centering
    \includegraphics[scale=0.4]{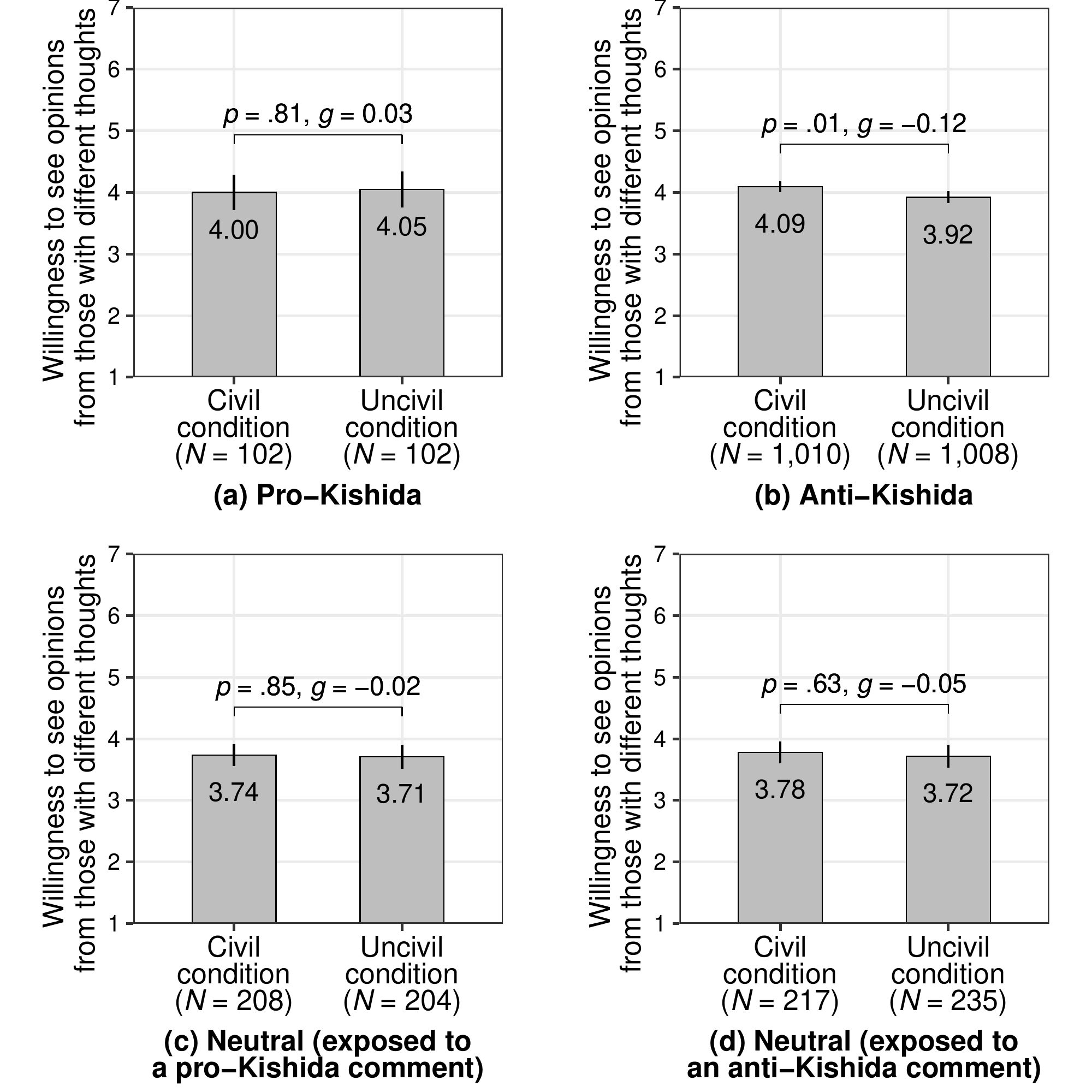}
    \caption{Results for Hypothesis 2}
    \label{figure_h2}
\end{figure}

Figure \ref{figure_h2} displays the results for Hypothesis 2. Again, the sizes of the bars represent the mean values of the outcome variable, and the lines on the bars are their 95\% confidence intervals.

Only the anti-Kishida participants (panel (b); upper right) show supporting results for Hypothesis 2. The anti-Kishida participants were less willing to see political opinions by people who have different thoughts than them on the internet when the presented comment was uncivil than when it was civil. The mean willingness of the anti-Kishida participants exposed to a civil comment was 4.09 with 95\% CI [4.00, 4.18], while the mean of those exposed to an uncivil comment was 3.92 with 95\% CI [3.82, 4.02]. The mean difference was -0.17 (Hedges' \textit{g} = -0.12, \textit{t}(1,972.95) = -2.57, \textit{p} = .0102).

On the contrary, the results of the other three subgroups do not provide supporting evidence for Hypothesis 2.
More specifically, for pro-Kishida participants (panel (a); upper left), the mean willingness of the participants exposed to a civil comment was 4.00 with 95\% CI [3.71, 4.29], while the mean of those exposed to an uncivil comment was 4.05 with 95\% CI [3.76, 4.34]. The mean difference between the conditions was 0.05 (Hedges' \textit{g} = 0.03, \textit{t}(200.03) = 0.24, \textit{p} = .81).
For neutral participants exposed to a pro-Kishida comment (panel (c); lower left), the mean willingness of the participants exposed to a civil comment was 3.74 with 95\% CI [3.56, 3.91], while the mean of those exposed to an uncivil comment was 3.71 with 95\% CI [3.52, 3.90]. The mean difference was -0.03 (Hedges' \textit{g} = -0.02, \textit{t}(400.63) = -0.19, \textit{p} = .85). 
For neutral participants exposed to an anti-Kishida comment (panel (d); lower right), the mean willingness of the participants exposed to a civil comment was 3.78 with 95\% CI [3.60, 3.96], while the mean of those exposed to an uncivil comment was 3.72 with 95\% CI [3.53, 3.90]. The mean difference was -0.06 (Hedges' \textit{g} = -0.05, \textit{t}(438.71) = -0.48, \textit{p} = .63).

The difference between the two conditions is statistically significant only in the anti-Kishida subgroup under multiple testing correction with the Benjamini-Hochberg method, and thus, the results support Hypothesis 2 only for the anti-Kishida subgroup.
The effect size in the anti-Kishida subgroup (Hedges' \textit{g} = -0.12) is relatively small, while statistically significant. However, it should be noted that this is the effect of just one comment; in real-world settings, people might be exposed to a large number of uncivil comments, and thus the effect might accumulate.

\section*{Discussion}
The present study hypothesized that individuals who are exposed to a political comment are (1) less willing to see other comments from the person who posted the comment and (2) less willing to see political comments from people who have different thoughts than them, when the comment is uncivil than when it is civil. To test these hypotheses, the present study conducted a preregistered online survey experiment targeting Japanese citizens, using the context of pro- vs anti-Kishida cabinet conflict. The results show that the participants were less willing to see other comments by the person who posted the comment when the comment was uncivil than when it was civil. In addition, the anti-Kishida participants were less willing to see political opinions by people with different thoughts than them when the comment was uncivil than when it was civil, while the participants in the other subgroups did not show a similar tendency.

These findings suggest that uncivil expressions in online political communication on social media might lead people to mute and unfollow those who post uncivil comments, and sometimes even avoid seeing opinions from those who have different thoughts than them. Given the previous study's finding that unfollowing, combined with social influence, promotes the formation of echo chambers on social media \citep{Sasahara2021}, the evidence obtained from the present study suggests that the proliferation of uncivil political expressions on social media might promote political echo chamber. In this sense, the present study provides deep insights into the mechanism behind the echo chamber phenomenon, which seems to be one of the key features of polarized politics.

Despite these contributions, the present study has several limitations.
Firstly, Hypothesis 2 was supported only for the anti-Kishida subgroup and not for the other three subgroups, and the reason for this heterogeneity remains unclear. This heterogeneity implies that the effect of incivility on individuals' willingness to see comments from those who have different thoughts than them might depend on the context regarding the issue that the comment mentions.
Second, there might be an issue of non-representative sample. The present study conducted an online survey with participants aged from 18 to 70. While a quota sampling approach was employed to obtain a sample that resembles the Japanese population structure, it should be carefully noted that those over 70 years old were not included in the sample. Furthermore, a selection bias might have been at play because those with higher political interest were more likely to participate in the survey.
Third, while the present study employed an RCT approach to provide robust causal evidence, it is not certain whether the findings of the present study can be generalized to other contexts and countries.
Therefore, further research is needed in which these limitations would be eliminated.

\theendnotes

\section*{Funding}
This study was financially supported by the JSPS KAKENHI Grant Number 22J21515, the Research Fund from the Quantitative Methods for International Studies Program at Kobe University, and the author's own resources.



\startappendix
\vspace{-7mm}
\subsection*{Manipulation Check Results}
\begin{figure}[ht]
    \centering
    \includegraphics[scale=0.4]{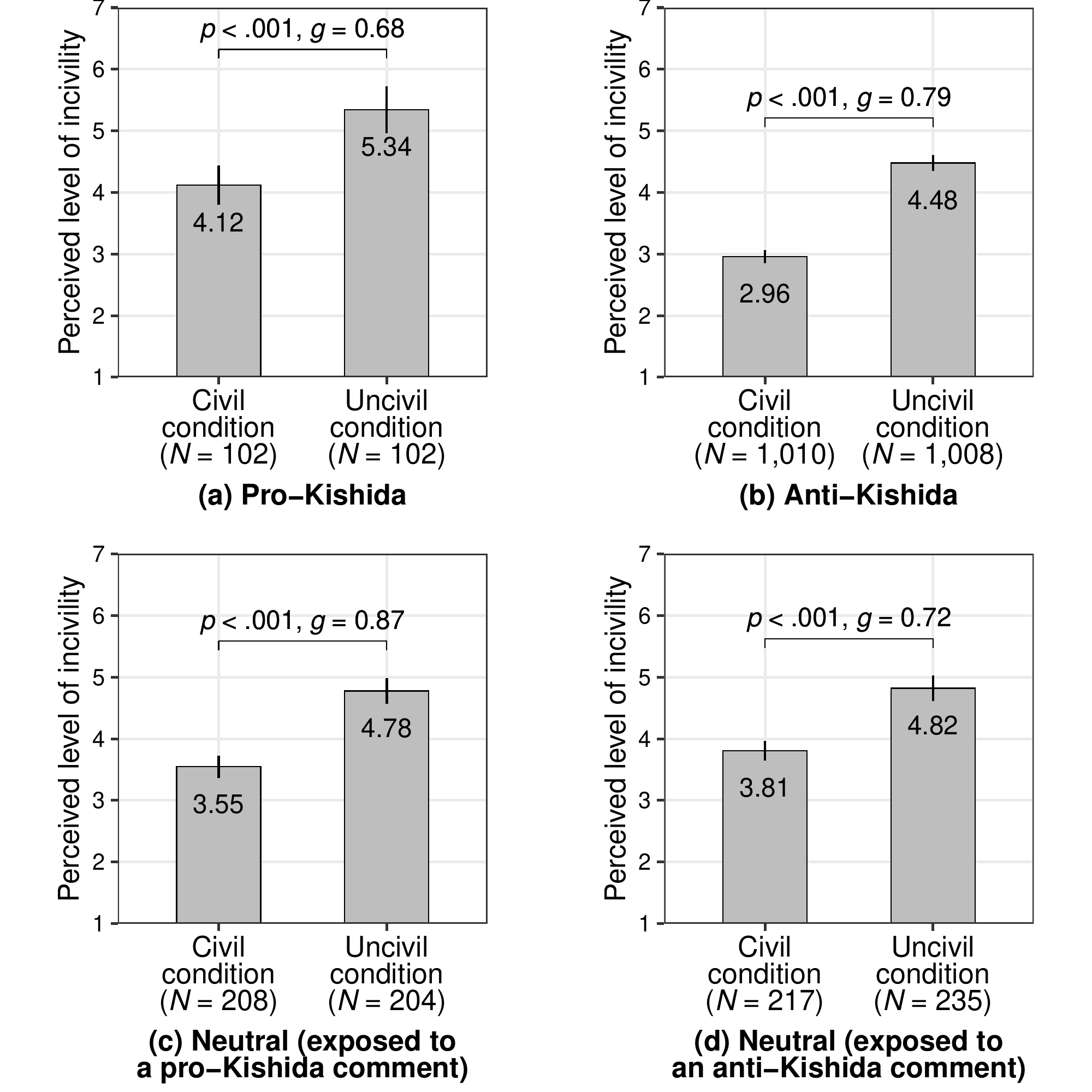}
    \caption{Manipulation Check Results}
    \label{figure_manipulation_check}
\end{figure}

\newpage
\subsection*{Balance Tables}
\begin{table}[h]
\centering
\caption{Balance of Pre-treatment Variables (Pooled)} 
\label{balance_all}
\begingroup\tablesize
\begin{tabularx}{125mm}{lRR}
  \hline
  & Civil condition & Uncivil condition \\ 
  \hline
Gender: Woman & 47.28\% & 48.01\% \\ 
  Gender: Man & 52.28\% & 51.74\% \\ 
  Gender: Others & 0.44\% & 0.25\% \\ 
  Age: 18-29 & 16.28\% & 15.48\% \\ 
  Age: 30-39 & 17.65\% & 17.09\% \\ 
  Age: 40-49 & 23.39\% & 23.44\% \\ 
  Age: 50-59 & 20.02\% & 21.96\% \\ 
  Age: 60-70 & 22.65\% & 22.02\% \\ 
  Education: Junior high school & 2.32\% & 2.17\% \\ 
  Education: High school & 26.43\% & 29.69\% \\ 
  Education: Junior college, etc. & 22.41\% & 21.30\% \\ 
  Education: College & 43.94\% & 41.37\% \\ 
  Education: Graduate school & 4.90\% & 5.47\% \\ 
  Income: Less than 2 million yen & 15.72\% & 17.44\% \\ 
  Income: 2-4 million yen & 23.80\% & 22.71\% \\ 
  Income: 4-6 million yen & 22.01\% & 21.32\% \\ 
  Income: 6-8 million yen & 15.42\% & 16.41\% \\ 
  Income: 8-10 million yen & 10.93\% & 10.26\% \\ 
  Income: 10 million yen or more & 12.13\% & 11.87\% \\ 
   \hline
\textit{N} & 1,603 & 1,621 \\ 
   \hline
\end{tabularx}
\endgroup
\end{table}

\newpage
\begin{table}[h]
\centering
\caption{Balance of Pre-treatment Variables (Pro-Kishida Participants)} 
\label{balance_pro_kishida}
\begingroup\tablesize
\begin{tabularx}{125mm}{lRR}
  \hline
  & Civil condition & Uncivil condition \\ 
  \hline
Gender: Woman & 40.20\% & 35.29\% \\ 
  Gender: Man & 59.80\% & 62.75\% \\ 
  Gender: Others & 0.00\% & 1.96\% \\ 
  Age: 18-29 & 14.71\% & 14.71\% \\ 
  Age: 30-39 & 13.73\% & 19.61\% \\ 
  Age: 40-49 & 29.41\% & 16.67\% \\ 
  Age: 50-59 & 16.67\% & 18.63\% \\ 
  Age: 60-70 & 25.49\% & 30.39\% \\ 
  Education: Junior high school & 1.98\% & 0.00\% \\ 
  Education: High school & 16.83\% & 32.67\% \\ 
  Education: Junior college, etc. & 23.76\% & 22.77\% \\ 
  Education: College & 53.47\% & 35.64\% \\ 
  Education: Graduate school & 3.96\% & 8.91\% \\ 
  Income: Less than 2 million yen & 21.18\% & 17.89\% \\ 
  Income: 2-4 million yen & 23.53\% & 17.89\% \\ 
  Income: 4-6 million yen & 24.71\% & 17.89\% \\ 
  Income: 6-8 million yen &  4.71\% & 15.79\% \\ 
  Income: 8-10 million yen & 7.06\% & 9.47\% \\ 
  Income: 10 million yen or more & 18.82\% & 21.05\% \\ 
   \hline
\textit{N} & 102 & 102 \\ 
   \hline
\end{tabularx}
\endgroup
\end{table}

\newpage
\begin{table}[h]
\centering
\caption{Balance of Pre-treatment Variables (Anti-Kishida Participants)} 
\label{balance_anti_kishida}
\begingroup\tablesize
\begin{tabularx}{125mm}{lRR}
  \hline
  & Civil condition & Uncivil condition \\ 
  \hline
Gender: Woman & 44.84\% & 44.27\% \\ 
  Gender: Man & 54.76\% & 55.53\% \\ 
  Gender: Others & 0.40\% & 0.20\% \\ 
  Age: 18-29 & 15.35\% & 14.88\% \\ 
  Age: 30-39 & 18.12\% & 16.27\% \\ 
  Age: 40-49 & 23.17\% & 24.11\% \\ 
  Age: 50-59 & 20.50\% & 21.33\% \\ 
  Age: 60-70 & 22.87\% & 23.41\% \\ 
  Education: Junior high school & 1.69\% & 2.20\% \\ 
  Education: High school & 26.42\% & 29.04\% \\ 
  Education: Junior college, etc. & 21.24\% & 20.56\% \\ 
  Education: College & 45.66\% & 42.91\% \\ 
  Education: Graduate school & 4.99\% & 5.29\% \\ 
  Income: Less than 2 million yen & 14.67\% & 16.20\% \\ 
  Income: 2-4 million yen & 23.36\% & 22.61\% \\ 
  Income: 4-6 million yen & 22.30\% & 22.26\% \\ 
  Income: 6-8 million yen & 16.55\% & 17.60\% \\ 
  Income: 8-10 million yen & 12.09\% & 10.14\% \\ 
  Income: 10 million yen or more & 11.03\% & 11.19\% \\ 
   \hline
\textit{N} & 1,010 & 1,008 \\ 
   \hline
\end{tabularx}
\endgroup
\end{table}

\newpage
\begin{table}[h]
\centering
\caption{Balance of Pre-treatment Variables (Neutral Participants Exposed to a Pro-Kishida Comment)} 
\label{balance_neutral_exposed_to_pro}
\begingroup\tablesize
\begin{tabularx}{125mm}{lRR}
  \hline
  & Civil condition & Uncivil condition \\ 
  \hline
Gender: Woman & 45.67\% & 56.93\% \\ 
  Gender: Man & 53.85\% & 43.07\% \\ 
  Gender: Others & 0.48\% & 0.00\% \\ 
  Age: 18-29 & 18.27\% & 13.73\% \\ 
  Age: 30-39 & 17.31\% & 19.12\% \\ 
  Age: 40-49 & 25.00\% & 25.49\% \\ 
  Age: 50-59 & 20.67\% & 23.04\% \\ 
  Age: 60-70 & 18.75\% & 18.63\% \\ 
  Education: Junior high school & 4.33\% & 2.94\% \\ 
  Education: High school & 26.44\% & 31.37\% \\ 
  Education: Junior college, etc. & 23.56\% & 23.53\% \\ 
  Education: College & 40.38\% & 36.27\% \\ 
  Education: Graduate school & 5.29\% & 5.88\% \\ 
  Income: Less than 2 million yen & 16.57\% & 19.76\% \\ 
  Income: 2-4 million yen & 25.14\% & 24.55\% \\ 
  Income: 4-6 million yen & 17.71\% & 21.56\% \\ 
  Income: 6-8 million yen & 16.57\% & 14.37\% \\ 
  Income: 8-10 million yen & 10.29\% & 10.78\% \\ 
  Income: 10 million yen or more & 13.71\% &  8.98\% \\ 
   \hline
\textit{N} & 208 & 204 \\ 
   \hline
\end{tabularx}
\endgroup
\end{table}

\newpage
\begin{table}[h]
\centering
\caption{Balance of Pre-treatment Variables (Neutral Participants Exposed to an Anti-Kishida Comment)} 
\label{balance_neutral_exposed_to_anti}
\begingroup\tablesize
\begin{tabularx}{125mm}{lRR}
  \hline
  & Civil condition & Uncivil condition \\ 
  \hline
Gender: Woman & 57.60\% & 54.89\% \\ 
  Gender: Man & 41.94\% & 45.11\% \\ 
  Gender: Others & 0.46\% & 0.00\% \\ 
  Age: 18-29 & 13.36\% & 17.02\% \\ 
  Age: 30-39 & 16.13\% & 16.60\% \\ 
  Age: 40-49 & 21.20\% & 18.72\% \\ 
  Age: 50-59 & 20.28\% & 26.38\% \\ 
  Age: 60-70 & 29.03\% & 21.28\% \\ 
  Education: Junior high school & 2.33\% & 0.85\% \\ 
  Education: High school & 28.37\% & 28.09\% \\ 
  Education: Junior college, etc. & 24.65\% & 21.70\% \\ 
  Education: College & 40.00\% & 44.26\% \\ 
  Education: Graduate school & 4.65\% & 5.11\% \\ 
  Income: Less than 2 million yen & 15.52\% & 20.10\% \\ 
  Income: 2-4 million yen & 23.56\% & 22.16\% \\ 
  Income: 4-6 million yen & 22.41\% & 17.53\% \\ 
  Income: 6-8 million yen & 17.24\% & 14.95\% \\ 
  Income: 8-10 million yen &  6.90\% & 11.34\% \\ 
  Income: 10 million yen or more & 14.37\% & 13.92\% \\ 
   \hline
\textit{N} & 217 & 235 \\ 
   \hline
\end{tabularx}
\endgroup
\end{table}

\end{document}